\documentclass[12pt]{article}

\usepackage{amsmath}
\usepackage{amsfonts}
\usepackage{amssymb}
\usepackage{latexsym}
\usepackage{color}
\usepackage[dvips]{graphicx}
\usepackage{cite}
\input{colordvi.tex}

\renewcommand{\thefootnote}{\#\arabic{footnote}}

\begin{document}
\setcounter{footnote}{0}

\begin{titlepage}
\begin{flushright}
RESCEU-43/12

\end{flushright}
\begin{center}


\vskip .5in

{\Large \bf
Constraining Primordial Magnetic Fields by CMB Photon-Graviton Conversion
}
\vskip .45in

{\large
Pisin Chen$^{1,2,3,4}$
and 
Teruaki Suyama$^{5}$
}

\vskip .45in%

{\em
$^1$
 Department of Physics, National Taiwan University, Taipei, Taiwan 10617\\
$^2$
 Graduate Institute of Astrophysics, National Taiwan University, Taipei, Taiwan 10617\\
$^3$
 Leung Center for Cosmology and Particle Astrophysics, National Taiwan University, Taipei, Taiwan 10617\\
$^4$
 Kavli Institute for Particle Astrophysics and Cosmology, SLAC National Accelerator Laboratory, Stanford University, Stanford, CA 94305, U.S.A.
  }\\
{\em
$^5$
   Research Center for the Early Universe (RESCEU), Graduate School
  of Science,\\ The University of Tokyo, Tokyo 113-0033, Japan
    }

\end{center}

\vskip .4in

\begin{abstract}
We revisit the method of using the photon-graviton conversion mechanism in the presence
of the external magnetic field to probe small-scale primordial magnetic fields
that may exist between the last scattering surface and present. Specifically, we investigate 
impacts on the conversion efficiency due to the presence of matter, including the plasma collective effect and the atomic polarizability. 
In general, these effects tend to reduce the conversion probability. Under this more realistic picture and based on the precision of COBE's measurement of CMB (cosmic microwave background) blackbody spectrum, we find an upper bound for the primordial magnetic field strength, $B\lesssim 30{\rm G}$, at the time of recombination. Although at present the bound based on the photon-graviton conversion mechanism is not as tight as that obtained by the direct use of CMB temperature anisotropy, it nevertheless provides an important independent constraint on primordial magnetic fields and at epochs in addition to the recombination. The bound can be significantly improved if the CMB blackbody spectrum measurement becomes more precise in future experiments such as PIXIE.

\end{abstract}
\end{titlepage}

\renewcommand{\thepage}{\arabic{page}}
\setcounter{page}{1}
\renewcommand{\thefootnote}{\#\arabic{footnote}}

\section{Introduction}
Magnetic fields in the universe have been observed at various scales 
(see, for instance, \cite{Grasso:2000wj,Widrow:2002ud} and references therein).
For instance, there exist intra-galactic magnetic fields at the level of ${\cal O}({\rm \mu G})$.
Magnetic fields also exist in cluster of galaxies.
However, their origin remains uncertain and even the possibility of
its primordial origin going back to inflation has been seriously considered \cite{
Turner:1987bw,Ratra:1991bn,Dolgov:1993vg,Lemoine:1995vj,Calzetta:1997ku,Davis:2000zp,Bamba:2003av,Bamba:2004cu,Martin:2007ue,Fujita:2012rb}.

In this equivocal situation, it is important to use as many independent methods 
as possible to probe magnetic fields on any scale and at any time to build 
up a consistent and robust picture of the evolution of the magnetic fields.
Among such probes, the standard one is to compute the temperature anisotropies
of the Cosmic Microwave Background (CMB) induced by the magnetic fields and to 
compare them with observations.
This method basically probes the magnetic field that might have existed prior
to recombination.
Assuming that the cosmic magnetic field strength scales as $B \propto a^{-4}$, due solely to the cosmic expansion,
where $a$ is the scale factor,
this places an upper bound on its present value 
as $B_{*}< {\cal O}(\rm {n G})$ (the pricise value depends on the assumption on the 
spectral tilt of the power spectrum of the magnetic fields, 
for instance, see \cite{Yamazaki:2012pg} and references therein).
Big Bang Nucleosynthesis(BBN) is another useful probe of the magnetic fields that
might have existed at the time of BBN (see for instance, \cite{Kawasaki:2012va}
and references therein).
The presence of the magnetic fields affects the expansion rate of the universe, which in turn changes  
the nuclear reaction rates and the abundance of light elements.
Observations of the light elements put the bound on the magnetic field as
$B_{*}< {\cal O}(\mu {\rm G})$ normalized by today's value.
High energy gamma-rays measured by High Energy Stereoscopic System (HESS)
and Fermi Telescopes recently derived lower bounds on the magnetic field (as a 
function of the coherent length scale of the magnetic fields) from the 
observation of TeV gamma-rays (by HESS) and non-observation of GeV gamma-rays 
(by Fermi Telescope), which is not compatible with zero magnetic fields \cite{Neronov:1900zz,Taylor:2011bn}.

Photon-graviton conversion is an alternative and completely different probe of the
magnetic field. 
The photon-graviton conversion, or in more general term the photon-graviton
oscillation, occurs whenever the photon propagates in the presence of the magnetic field.
The background magnetic field induces a coupling between photon and graviton states, which causes the mixing of propagation eigenstates.
This phenomenon itself was first noticed by Gertsenshtein \cite{Gertsenshtein:1962}.
One of the present authors (P.S.) invoked this mechanism to investigate cosmological magnetic fields 
by using the CMB as the incident photons \cite{Chen:1994ch}.
The magnetic fields, if they do exist after the recombination, would convert some of the
CMB photons into gravitons which are not detectable, results in the reduction of CMB 
intensity from the blackbody radiation spectrum.
How much the CMB intensity is diminished depends on the magnetic field strength.
In \cite{Chen:1994ch}, it was pointed out that the detection/non-detection of the deviation
of the CMB distribution from the perfect blackbody spectrum should enable us to detect/constrain the
strength of the magnetic field. 

This paper aims at refining the calculation of the probability of the 
photon-graviton conversion for the CMB photons derived in \cite{Chen:1994ch} by
taking into consideration two matter effects: plasma oscillations and atomic polarizability, 
so as to deduce a more reliable constraint on primordial magnetic fields.
While the effects due to atomic polarizability was briefly discussed \cite{Chen:1994ch}, 
that due to plasma oscillations was not taken into account in the original paper.
Such matter effects tend to significantly change (in general suppress) the conversion
probability compared to that in the vacuum case (for the effect of the plasma oscillation alone,
see \cite{Cillis:1996qy}).
We will find that the current bound on the deviation of the CMB from the blackbody
radiation places the upper bound on the magnetic field strength normalized
by today's value at about $30~{\rm G}$ or larger, depending on the coherent
length of the magnetic field.
This bound can be improved by using the proposed future experiments such as PIXIE \cite{Kogut:2011xw}.

\section{Photon-graviton conversion}
In the presence of background magnetic field,
the electromagnetic wave and gravitational wave are coupled with each other,
which results in the conversion between them.
Considering plane waves traveling along the $z$-axis, 
we can construct two convenient variables $H_1$ and $H_2$, defined by
\begin{equation}
H_1=\frac{1}{\sqrt{8\pi G}} \left( \sin \beta h_+-\cos \beta h_\times \right),~~~~~
H_2=\frac{1}{\sqrt{8\pi G}} \left( \cos \beta h_++\sin \beta h_\times \right),
\end{equation}
where $h_+$ and $h_\times$ are standard polarization modes of gravitational waves,
$\beta$ is defined by ${\vec B}_\bot=B_\bot(\cos \beta, \sin \beta)$ and
${\vec B}_\bot$ is the magnetic field projected onto the $x-y$ plane.
Linearized Einstein equations and Maxwell equations show that wave equations
are separable into two decoupled set of equations that take exactly the same form: 
a set of two equations that contains only $H_1$ and $A_x$ 
($x$-component of the vector potential) and the other set that contains only $H_2$ and $A_y$.
Explicit forms of those equations are given by
\begin{equation}
\left(
  \begin{array}{cc}
    \omega+i\partial_z & 0 \\
    0 & \omega+i\partial_z+\Delta_A
  \end{array}
  \right)
  \left( \begin{array}{cc} H\\ A\\ \end{array} \right)=
  \sqrt{8\pi G}B_\bot
  \left(
  \begin{array}{cc}
    0 & -i \\
    i & 0
  \end{array}
  \right)
  \left( \begin{array}{cc} H\\ A\\ \end{array} \right). \label{eq-mixing}
\end{equation}
Here $H$ is either $H_1$ or $H_2$ and $A$ is the corresponding vector potential.
The term $\Delta_A$, whose explicit expression will be provided later, 
represents the modification of the index of refraction due to matter effects.
(matter effects for gravitational wave are neglected due to its extreme smallness in realistic situations).

Assuming the wavelength of the plane waves 
(we will eventually apply this formalism to the case where the electromagnetic wave is CMB) 
is much shorter than the length scale associated with the variation of the background magnetic field,
which is actually satisfied for a wide range of scales below the CMB scales 
({\it i.~e.}, $\sim {\rm Mpc}$),
we can integrate Eq.~(\ref{eq-mixing}) to obtain the solution expanded
up to first order in $\Delta_M \equiv \sqrt{8\pi G}B_\bot$ as
\begin{eqnarray}
&&H(z)=e^{i\omega z}
\bigg[ H(0)-A(0) \int_0^z dz'~\Delta_M(z') e^{i\int_0^{z'}dz'' \Delta_A(z'')} \bigg], \\
&&A(z)=e^{i\omega z+i \int_0^z dz' \Delta_A(z')}
\bigg[ A(0)+H(0) \int_0^z dz'~\Delta_M(z') e^{-i\int_0^{z'}dz'' \Delta_A(z'')} \bigg].
\end{eqnarray}
These equations clearly show that $H$ and $A$ convert into each other (although the full conversion
cannot be treated by the leading-order expansion in $\Delta_M$).
In terms of the quantum mechanical language,
this result can be interpreted as conversion between photons and gravitons.

We are interested in a situation where electromagnetic wave,
which is the only existing component initially, converts into
gravitational wave during propagation.
In such a case, the conversion probability from photon to graviton during
propagation over a distance $d$ is given by
\begin{equation}
P(\gamma \to g)=1-\frac{{|A(d)|}^2}{{|A(0)|}^2}={\bigg| \int_0^d d\ell~\Delta_M(\ell) 
\exp \left( i \int_0^\ell d\ell'~\Delta_A (\ell') \right) \bigg|}^2,
\label{conversion-p}
\end{equation}
where we have used $d$ and $\ell$ instead of $z$ in order to avoid confusion
with redshift parameter $z$ that will be used later.
For $B_\bot$ close to some critical value, 
for which Eq.~(\ref{conversion-p}) approaches unity, 
the higher order terms in $\Delta_M$ must be taken into account.
This means Eq.~(\ref{conversion-p}), which we rely on, is applicable only for magnetic
fields smaller than the critical value.
Nonetheless, we can still obtain meaningful constraint on
the amplitude of the magnetic field within the domain of Eq.~(\ref{conversion-p}).
This is basically due to our use of CMB which is experimentally confirmed 
to obey the Planck distribution down to ${\cal O}(10^{-4})$.
Photon-graviton conversion with probability larger than ${\cal O}(10^{-4})$
would result in the deviation of the CMB spectrum from the Planck distribution, which would be
inconsistent with existing measurements.
We will come back to this point in the next section.

In principle, we can perform double integral of Eq.~(\ref{conversion-p}) numerically
to compute the conversion probability.
This can be done straightforwardly, but takes some computation time.
Instead of taking this approach, here we make an approximation for Eq.~(\ref{conversion-p})
to analytically perform the integration by treating the phase $\int_0^\ell d\ell'~\Delta_A (\ell')$
as a huge ({\it i.~e.}, its absolute value is much bigger than unity) 
and rapidly changing quantity over distance, except for some points
where $\Delta_A$ becomes accidentally zero if that ever happens.
Assuming that there do exist some sites where $\Delta_A$ becomes zero,
which is not valid for some frequency range of the electromagnetic wave,
we can apply the saddle point method to perform the integral of Eq.~(\ref{conversion-p})
and we end up with an expression given by
\begin{equation}
P(\gamma \to g) \approx 2\pi \sum_{d_A} \Delta_M^2(d_A) {\bigg| \frac{d\Delta_A(\ell)}{d\ell} \bigg|}_{\ell=d_A}^{-1}, \label{conversion-p2}
\end{equation}
where $d_A$ is the position in which $\Delta_A(d_A)=0$.
Only the small region with size given by ${|d\Delta_A/d\ell|}^{-1/2}$ surrounding the
point where $\Delta_A=0$ contributes to the conversion probability.

Eq.~(\ref{conversion-p2}) has been derived under the assumption that the
magnetic field is uniform both in magnitude and direction within the 
domain where most of the conversion occurs.
If the coherent length $\ell_c$ of the magnetic field is shorter than the size of 
the domain, the conversion still occurs within the domain, but proceeds
independently in each small region where magnetic field can be regarded as uniform.
Therefore, the total conversion probability becomes the sum of conversion
probabilities evaluated for each small region with its size given by $\ell_c$.
Then, for $\ell_c < R_A$, Eq.~(\ref{conversion-p2}) is modified into
\begin{equation}
P(\gamma \to g) \approx 2\pi \sum_{d_A} \Delta_M^2(d_A) {\bigg| \frac{d\Delta_A(\ell)}{d\ell} \bigg|}_{\ell=d_A}^{-1} \times \left( \frac{\ell_c}{R_A} \right), \label{conversion-p3}
\end{equation}
where $R_A \equiv {|d\Delta_A/d\ell|}^{-1/2}$.
Therefore, the conversion probability is suppressed by a factor $\ell_c/R_A$ for $\ell_c < R_A$,
relative to the case where $\ell_c > R_A$. 

\begin{figure}[t]
  \begin{center}{
    \includegraphics[scale=0.9]{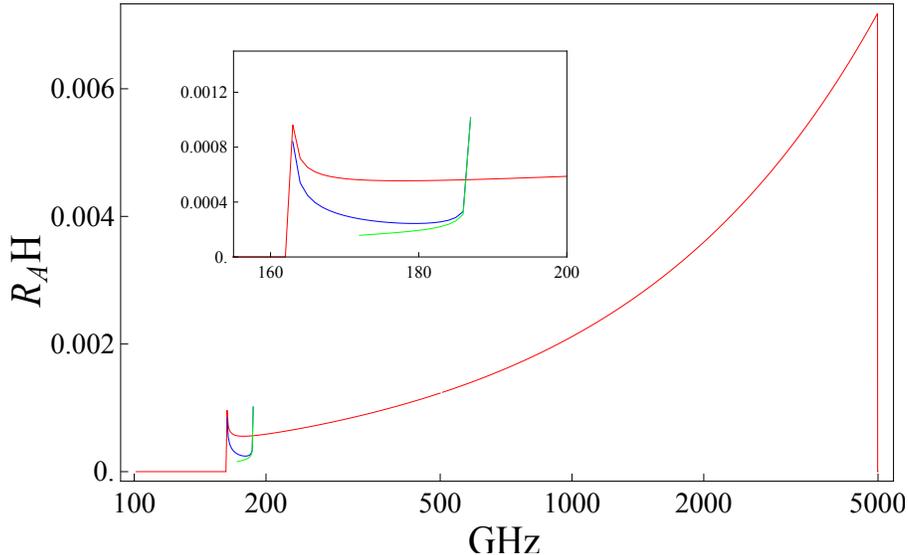}
    }
  \end{center}
  \caption{Ratio of $R_A$ to the horizon length as a function of
  the photon frequency measured today.
  }
 \label{cp3}
\end{figure}

In Fig.~\ref{cp3}, we show the quantity $R_A H$, which measures the ratio
of $R_A$ to the horizon length at the time when $\Delta_A=0$, as a function
of today's photon frequency.
(For the evaluation of $\Delta_A$, see the next subsection).
For the frequency interval $162~{\rm GHz} < \nu < 187~{\rm GHz}$,
$R_A H$ becomes multivalued due to the existence of multiple solutions of 
$\Delta_A=0$.
We find that for $100~{\rm GHz}< \nu < 10^3~{\rm GHz}$, 
which is the main frequency range of the CMB measurement,
$R_A H< 2\times 10^{-3}$.
Thus, as a conservative estimate, the boundary at which $\ell_c=R_A$ 
is given by $\ell_c = 2\times 10^{-3}H^{-1}$. 

\subsection{Matter effects}
Various non-vacuum effects enter $\Delta_A$ that modifies the dispersion
relation of electromagnetic wave in vacuum.
In this paper, we take into account two matter effects,
{\it i.~e.}, plasma frequency and atomic polarizability. 
While the effects due to atomic polarizability was briefly discussed \cite{Chen:1994ch}, 
that due to plasma oscillations was not taken into account in the original paper.

For an electromagnetic wave with frequency $\omega$ and wave number $k$ propagating in a plasma, 
its dispersion relation is modified into
\begin{equation}
\omega^2=k^2+\omega_p^2,~~~~~\omega_p^2=\frac{4\pi \alpha n_e}{m_e},
\end{equation}
where $\alpha \simeq 1/137$ is the fine structure constant,
$n_e$ is the number density of free electrons, $m_e$ is the electron mass, and $\omega_p$ is the plasma frequency.
As a result, the collective oscillations of plasma contribution to $\Delta_A$ is given by
\begin{equation}
\Delta_{A1}=-\frac{1}{2}\frac{\omega_p^2}{\omega}=-\frac{2\pi \alpha n_e}{m_e \omega}, \label{plasma}
\end{equation}
which is frequency dependent.

We next consider the atomic polarizability,
which also modifies the propagation of the electromagnetic wave from the vacuum case.
In the presence of external electric field ${\vec E}$,
atomic gas acquires electric polarization ${\vec P}$ given by
\begin{equation}
{\vec P}=\alpha_p {\vec E},
\end{equation}
where $\alpha_A$ is the polarizability of atom $A$.
This produces the electric permittivity $\epsilon$ given by
\begin{equation}
\epsilon=\epsilon_0+n_H \alpha_H, \label{permittivity}
\end{equation}
where $n_H$ is the number density of neutral hydrogen atoms\footnote{
Only in this section, we work in the SI unit.}.
The helium gas also contributes to $\epsilon$ through the combination 
$n_{H_e} \alpha_{H_e}$. 
In the cosmological background, $n_{H_e}/n_H \simeq 1/12$.
Theoretical calculations give each of the polarizability as\cite{Handbook}  
\begin{equation}
\frac{\alpha_H}{4\pi \epsilon_0}=0.667\times 10^{-30}~{\rm m}^{-3},~~~~~
\frac{\alpha_{H_e}}{4\pi \epsilon_0}=0.205\times 10^{-30}~{\rm m}^{-3}. \label{pol-alpha}
\end{equation}
Using these values, we find that the ratio of helium to hydrogen contributions is
given by
$(n_{H_e}\alpha_{H_e})/(n_H \alpha_H) \simeq 0.026$, which is quite tiny.
We therefore neglect the helium contribution.
Using Eq.~(\ref{permittivity}), $\Delta_A$ contributed by the atomic
polarizability is given by
\begin{equation}
\Delta_{A2}=\frac{n_H \alpha_H \omega}{2\epsilon_0}. \label{polarizability}
\end{equation}
Contrary to the plasma frequency, the atomic polarizability yields positive
contribution to $\Delta_A$.
Notice that the atomic polarizability $\alpha_H$ in Eq.~(\ref{pol-alpha}) 
is valid only for $\nu < \nu_{\rm excite}=2.44\times 10^6~{\rm GHz}$, 
corresponding to the energy difference between the ground state and the excited state of the hydrogen atom.
The relative error caused by the use of Eq.~(\ref{pol-alpha}) scales as $\sim {(\nu/\nu_{\rm excite})}^2$,
which becomes ${\cal O}(1)$ at the time of recombination for the CMB photons with current frequency $10^3~{\rm GHz}$.
However, as Fig~\ref{cp1} shows, all the CMB frequencies evaluated at the redshifts when
$\Delta_A$ vanishes are considerably lower than $\nu_{\rm excite}$ (the corresponding error is at most $1~\%$),
which justifies the use of Eq.~(\ref{polarizability}).

Total $\Delta_A$ is then given by the sum of Eq.~(\ref{plasma}) and Eq.~(\ref{polarizability});
\begin{eqnarray}
\Delta_A&=&-\frac{2\pi \alpha n_e}{m_e \omega} \left( 
1-\frac{\alpha_H}{4\pi \epsilon_0} \frac{m_e \omega^2}{\alpha} \frac{1-X_e}{X_e} \right), \nonumber \\
&=&-\frac{\omega^2_p}{2\omega} 
\left( 1-0.0061~{\left( \frac{\omega}{1~{\rm eV}} \right)}^2 \frac{1-X_e}{X_e} \right), \label{Delta_A}
\end{eqnarray}
where $X_e$ is the ionization fraction \footnote{
The coefficient $0.0061$ appearing in Eq.~(\ref{Delta_A}) is different by about
$15~\%$ from $0.0073$ given in \cite{Mirizzi:2009iz, Mirizzi:2009nq}, where the polarizability of hydrogen molecules, $\alpha_{H^2}/(4\pi \epsilon_0)\simeq 0.80 \times 10^{-30}~m^{-3}$\cite{Handbook}, was apparently used. 
}.
In the cosmological background, both the frequency of the electromagnetic wave
and the ionization fraction change due to cosmic expansion.
Thus, it is possible for a wide range of $\omega$ that the quantity in the 
parenthesis of Eq.~(\ref{Delta_A}) becomes zero at some particular time (this depends on $\omega$) 
between the last scattering surface and present time.
The point is that we do not need to fine-tune the frequency to have vanishing $\Delta_A$.
Vanishing of $\Delta_A$ is automatically achieved for a wide range of $\omega$ 
in the course of cosmic expansion.

The redshift dependence of the plasma frequency $\omega_p$ is given by
\begin{equation}
\omega_p (z)=1.5 \times 10^{-14}~{\rm eV} ~{(1+z)}^{3/2} X_e^{1/2}(z).
\end{equation}
As for the ionization fraction, we compute it by numerically solving the
evolution equation given in \cite{weinberg:cosmology}.

\section{Result}
We numerically evaluate the conversion probability from photon to graviton given
by Eq.~(\ref{conversion-p2}) with the use of Eq.~(\ref{Delta_A}) as a function
of the frequency of photon.
With CMB as the propagating photons in mind, the frequency range we consider is from 
$100~{\rm GHz}$ to $6~{\rm THz}$.
We fix the amplitude of the magnetic field at the time of last scattering to be
$1~{\rm G}$ and assume that it scales as ${(1+z)}^2$ based on the conservation of magnetic flux 
and due to the cosmic expansion.
To compute the conversion probability for different values of the magnetic field,
we simply need to rescale it by the square of the magnetic field in the unit of $1~{\rm G}$ 
because the conversion probability is just proportional to $B^2$.
If there is no solution of $\Delta_A(z)=0$ in the redshift interval
$z_{\rm re}=11.4<z<z_{\rm rec}=1090$, we assign zero to the conversion probability.
Strictly speaking, even if $\Delta_A=0$ is not achieved along the photon path,
the integral of Eq.~(\ref{conversion-p}) does not vanish exactly in general.
But the conversion probability for such a case is expected to be highly suppressed 
due to the efficient phase cancellation compared to the one for which $\Delta_A=0$ 
is satisfied at some time.
The lower end $z_{\rm re}=11.4$ corresponds to the reionization epoch\cite{Ade:2013zuv}.
The reionization invalidates, for $z<z_{\rm re}$, 
the use of $X_e$ obtained under the approximation that only the 
cosmological recombination is relevant to the ionization fraction.
Since the quantitative estimation of $\Delta_A$ after the reionization
is beyond the scope of our paper because of complexity of astrophysical processes, 
in order to be conservative as possible,
we simply do not take into account the  photon-graviton conversion that 
may happen after the reionization.

\begin{figure}[t]
  \begin{center}{
    \includegraphics[scale=1.0]{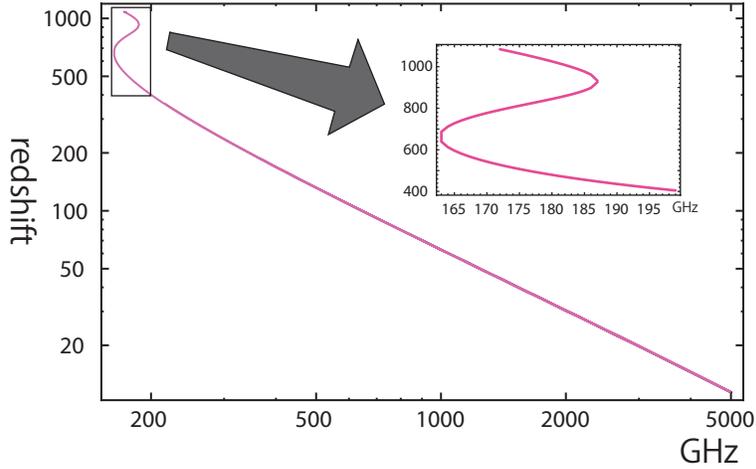}
    }
  \end{center}
  \caption{Redshift at which $\Delta_A$ becomes zero as a function of the photon frequency
  measured today. For the frequency interval $162~{\rm GHz} < \nu < 187~{\rm GHz}$,
  there are multiple such redshifts.}
 \label{cp1}
\end{figure}

Fig.~\ref{cp1} shows the redshift at which $\Delta_A$ becomes zero 
as a function of the photon frequency measured today.
The curve in Fig.~\ref{cp1} exhibits somewhat complicated behavior. 
As we see, there is no solution of $\Delta_A(z)=0$ for $\nu<162~{\rm GHz}$.
For $162~{\rm GHz} < \nu < 172~{\rm GHz}$, there are two solutions both of
which are between $500$ and $800$.
Between $172~{\rm GHz}$ and $187~{\rm GHz}$, three solutions exist.
Above $187~{\rm GHz}$, there is only one solution which monotonically
decreases as we increase the frequency and the redshift becomes
as small as $10$ for $\nu \simeq 5000~{\rm GHz}$.

\begin{figure}[t]
  \begin{center}{
    \includegraphics[scale=0.8]{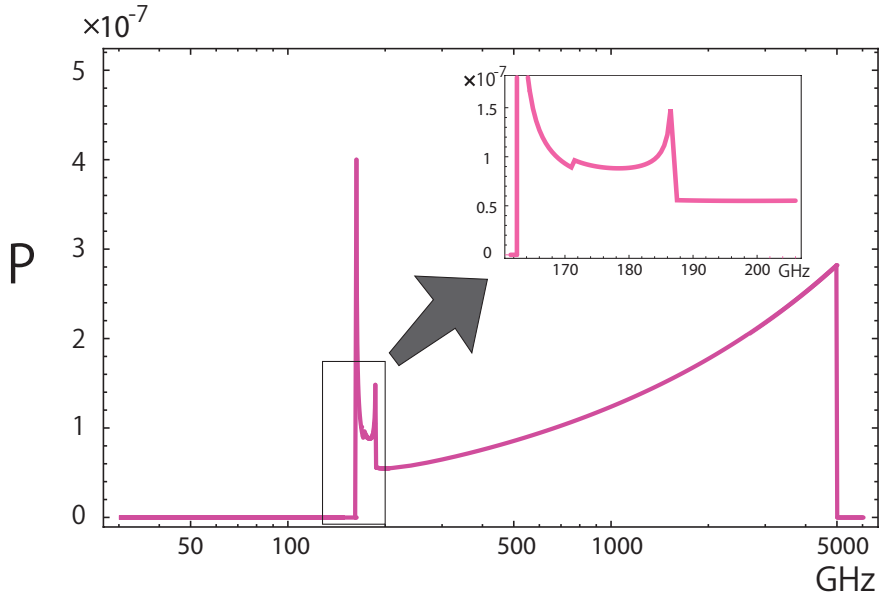}
    }
  \end{center}
  \caption{The photon-graviton conversion probability as a function of photon
  frequency today. Magnetic field of $1~{\rm G}$ at the time of recombination is assumed.
  }
 \label{cp2}
\end{figure}

Fig.~\ref{cp2} shows the conversion probability from photon to graviton
as a function of the photon frequency measured today assuming magnetic field
of $1~{\rm G}$ at the time of recombination.
Because of the absence of the solution of $\Delta_A=0$ for $\nu<162~{\rm GHz}$,
the conversion probability is set to be zero for that frequency range as 
mentioned above.
Just above $162~{\rm GHz}$, the conversion probability suddenly increases
up to $4\times 10^{-7}$ and gradually decays down to $10^{-7}$ until $172~{\rm GHz}$.
Above this frequency, number of solutions of $\Delta_A=0$ increases by one,
which results in a slight enhancement of the conversion probability across
this frequency.
From $172~{\rm GHz}$ to $187~{\rm GHz}$, the conversion probability mildly 
increases.
At $\nu=187~{\rm GHz}$, the number of solution of $\Delta_A=0$
becomes one and, as a result, conversion probability suddenly drops by a factor of $3$.
Above this frequency, the conversion probability monotonically grows but
very mildly.
Even at $\nu=5~{\rm THz}$, the conversion probability differs from the one
at $\nu=187~{\rm GHz}$ only by a factor of $6$.
As we have just found, the detailed behavior of the conversion probability
is highly complicated as a function of $\nu$, but its frequency dependence
is not so significant.
Therefore, we can crudely say that the conversion probability above the
critical frequency $\nu \sim 160~{\rm GHz}$ is a few times $10^{-7}$
for the magnetic field of $1~{\rm G}$ at the time of recombination
(up to $\nu \simeq 5~{\rm THz}$).

The photon-graviton conversion mechanism applied to the CMB photons
results in the frequency dependent modification of the distribution function 
from the Planck one.
Far Infrared Absolute Spectrophotometer (FIRAS) on board of the Cosmic
Background Explorer (COBE) confirmed for the frequency interval 
$60~{\rm GHz} < \nu < 600~{\rm GHz}$ that the CMB distribution is consistent
with the blackbody form within $\sim 10^{-4}$.
Using the scaling $P \propto B^2$ and crude approximation that 
$P \simeq 10^{-7}$ for $B=1~{\rm G}$, 
the FIRAS constraint on the deviation of the CMB spectrum yields the upper 
bound on the magnetic field strength as $B \lesssim 30~{\rm G}$
at the time of recombination.
This bound is weaker than the one obtained by the direct use of the CMB
temperature anisotropies, which yields $B < {\cal O}({\rm 10^{-3}~G})$ 
at the time of recombination \cite{Chen:1994ch}.
Nevertheless, the bound from the photon-graviton conversion is important in
that it provides the independent constraint from the CMB one by using the 
completely different mechanism and that the photon-graviton conversion can probe
the magnetic fields at epochs and scales not covered by the CMB temperature anisotropies.

\begin{figure}[t]
  \begin{center}{
    \includegraphics[scale=0.8]{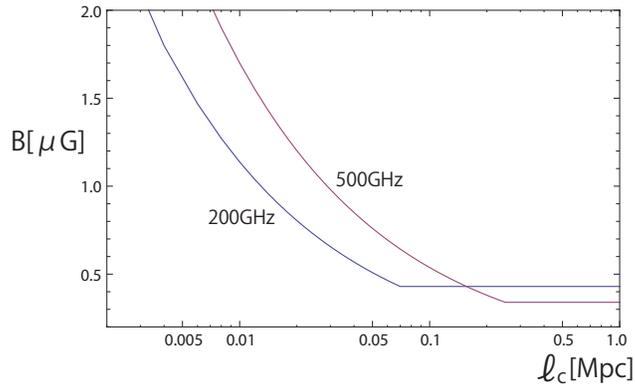}
    }
  \end{center}
  \caption{Magnetic field strength today that yields the photon-graviton
  conversion probability of $10^{-8}$ as a function of the magnetic 
  coherent length $\ell_c$ measured today.
  }
 \label{cp4}
\end{figure}

\section{Summary}

So far, our argument has been based on Eq.~(\ref{conversion-p2}).
As we have discussed in the last section, the conversion probability should be modified
and replaced by Eq.~(\ref{conversion-p2}) if the coherent length of the magnetic 
field becomes smaller than the critical value $R_A$ which depends on the photon 
frequency.
Fig.~\ref{cp4} shows the magnetic field strength today as a function of the
coherent length today required to yield the conversion probability of $10^{-8}$
for two photon frequencies $\nu=200~{\rm GHz}$ and $500~{\rm GHz}$.
This level of deviation from the Planck distribution is far below the present 
sensitivity, which is $\sim 10^{-4}$, 
but is expected to be achieved in future experiments such as PIXIE \cite{Kogut:2011xw}.
We find that the bound on the magnetic field strength for $\ell_c > R_A$ is 
$\lesssim 0.4~{\rm \mu G}$ (the precise value depends on the photon frequency).
This value is comparable to the BBN bound.
For $\ell_c < R_A$, the bound on $B$ scales as $B \propto \ell_c^{-1/2}$.
The value of $R_A$ depends on the photon frequency but, for the range of our interest,
it is ${\cal O}(0.1~{\rm Mpc})$.
Some literature \cite{Jedamzik:1996wp,Jedamzik:1996wp} suggest the magnetic damping
becomes non-negligible below this scale and our simple scaling of the magnetic field
$B \propto a^{-4}$ would accordingly be modified for such a case.\\

{\bf Acknowledgments:} 
TS thanks the Leung Center for Cosmology and Particle Astrophysics (LeCosPA), 
National Taiwan University for the kind hospitality during his visit when this project initiated.
This work is supported by Grant-in-Aid for Scientific Research on Innovative Areas
No.~25103505 (TS) from The Ministry of Education, Culture, Sports, Science and Technology (MEXT).

\bibliographystyle{unsrt}
\bibliography{draft}

\end{document}